\documentstyle[amssymb,aps]{revtex}

\begin{document}
\title{Trapped ions in laser fields: a benchmark for deformed quantum oscillators}
\author{V. Man'ko$^{\dagger \dagger }$, G. Marmo$^{\ddagger }$, A. Porzio$^{\dagger
} $, S. Solimeno$^{\dagger }$, and F. Zaccaria$^{\ddagger }$}
\address{Dip. di Fisica, Univ. ''Federico II'', Napoli \\
and \\
$\dagger $ Istituto Nazionale Fisica della Materia, Unit\`{a} di Napoli\\
$\ddagger $Istituto Nazionale Fisica Nucleare, Sez. Napoli\\
$^{\dagger \dagger }$Lebedev Physics Institute, Moscow}
\maketitle

\begin{abstract}
Some properties of the non--linear coherent states (NCS), recognized by
Vogel and de Matos Filho as dark states of a trapped ion, are extended to
NCS on a circle, for which the Wigner functions are presented. These states
are obtained by applying a suitable displacement operator $D_{h}\left(
\alpha \right) $ to the vacuum state. The unity resolutions in terms of the
projectors $\left| \alpha ,h\right\rangle \left\langle \alpha ,h^{-1}\right|
,\left| \alpha ,h^{-1}\right\rangle \left\langle \alpha ,h\right| $ are
presented together with a measure allowing a resolution in terms of $\left|
\alpha ,h\right\rangle \left\langle \alpha ,h\right| $. $D_{h}\left( \alpha
\right) $ is also used for introducing the probability distribution funtion $%
\rho _{A,h}\left( z\right) $ while the existence of a measure is exploited
for extending the P-representation to these states. The weight of the n-th
Fock state of the NCS relative to a trapped ion with Lamb-Dicke parameter $%
\eta ,$ oscillates so wildly as $n$ grows up to infinity that the normalized
NCS fill the open circle $\eta ^{-1}$ in the complex $\alpha $-plane. In
addition this prevents the existence of a measure including normalizable
states only. This difficulty is overcome by introducing a family of
deformations which are rational functions of n, each of them admitting a
measure. By increasing the degree of these rational approximations the
deformation of a trapped ion can be approximated with any degree of accuracy
and the formalism of the P-representation can be applied.
\end{abstract}

\section{\bf Introduction}

The theory of certain one-parameter (q--) deformations of Lie algebras, the
so called quantum groups has been of great interest in the last decade in
several areas of physics. In 1989 Biedenharn [1] and McFarlane [2]
independently defined the q-analogue coherent state of a deformed
q-oscillator, for which Nelson et al. [3] were able to obtain the resolution
of unity. Since then the properties of a class of deformations of the
harmonic oscillator were considered by several authors (see f.i. [4]).
Deformed quantum oscillators are represented by dynamical variables $A$, $%
A^{\dagger }$ and $N_{A}$ satisfying the commutation relations, $%
[A,N_{A}]=A, $ $[A^{\dagger },N_{A}]=-A^{\dagger }$ and $[A,A^{\dagger
}]=f\left( N_{A}\right) $, with $f\left( N_{A}\right) $ an arbitrary real
function of $N_{A}$. All such variables are constructed in terms of
single-mode field operators\footnote{%
wherever possible operators will be indicated by simple letters, except for
the addition of a caret when confusion could arise with c-number quantities} 
$a,\ a^{\dagger }$ and $a^{\dagger }a.$

In 1993 Crnugelj et al. [5] observed that the multiphoton interaction of a
single mode laser field with a two level atom is described by
deformed-oscillator creation and annihilation operators which in combination
with the pseudo-spin atomic operators $\sigma _{+}$ and $\sigma _{-}$, form
the potential 
\[
W_{JC}=A^{\dagger }\sigma _{-}+A\sigma _{+} 
\]
used in the Jaynes-Cummings model (JCM) [6].

In that period the J-C model was at center of the attention for the study of
laser cooling of ions placed in parabolic traps, with the quantized
center-of-mass motion of the ion playing the role of the boson mode, coupled
via the laser to the internal degrees of freedom. In the case of cooling the
operator $A$ is represented by a combination of some power of the
annihilation operator $a$ times a function of $n$. When some bosons of the
oscillator mode are destroyed the ion is excited to the upper level from
where it decays radiatively. Cooling was investigated in Lamb-Dicke [7] and
strong-sideband [8] limits, that is for ion excursions small compared with
the radiation wavelength. This study led to the discovery of many intriguing
effects connected with the nonclassical properties of the field, like as a
long-time sensitivity to the statistical properties of the radiation field
[9]. For example, the mean excitation number of the quantized oscillations
of a ion driven by a squeezed field exhibited periodic collapses and
revivals [10].

The interest for the vibrational motion of trapped ions was also motivated
by the connection between the state of motion and the properties of the
fluorescence spectra [11,12]. This link led some experimentalists to look
for new non classical radiation states generated by trapped ions forced into
some unusual vibrational states. In analogy to the preparation of
nonclassical states of light in quantum optics several authors examined the
preparation of the center-of-mass motion in a quantum state having no
classical counterpart. Worthy examples were those of Cirac et al. [13] who
considered the possibility of generating squeezed states of the vibrational
motion by irradiating the trapped ion with two standing-wave light fields of
different frequencies and locating the center of the trap potential at a
common node of both waves. In all these cases the nonlinear dependence of $A$
on $a,a^{\dagger }$ and $\hat{n}$, stemmed from the ion motion in the trap
potential.

de Matos Filho and Vogel [14] observed in 1993 that the center-of-mass state
of a trapped ion driven by a two-mode laser field decays toward a dark state
coincident with a nonlinear coherent state (hereinafter called NCS) of a
deformed oscillator. This result brought new fuel to the study of deformed
oscillators describing different classes of states arising in the trapped
ion motion under the action of two or three fields detuned by multiples of
the vibrational frequency (see f.i. [15] for nonlinear cat states). In the
wake of this interest attention was paid to theoretical models of deformed
oscillators, like those connected with excited coherent states and binomial
states [16]. All these non-linear oscillators differ for the deformation
function $h\left( \hat{n}\right) $ connecting the annihilation operator $a$
to the deformed one $A=ah\left( \hat{n}\right) $. The ancestor of these
realizations were the q-oscillators characterized by a deformation $%
h_{q}\left( \hat{n}\right) =\sqrt{\frac{\sinh \left( \lambda \hat{n}\right) 
}{\hat{n}\sinh \lambda }}$ increasing with $n$. Contrarily the trapped ion
deformation is a very irregular function of $n$, taking positive and
negative values. What is worse, for some combinations of the Lamb-Dicke
parameter $\eta ^{2}$ and $n$ it can vanish or become infinite. As a
consequence it is hard to capitalize on the work done for the q-oscillator
for studying the NCS of a trapped ion. In particular, while for the q-case
it has been found a measure resolving the unity, the same is not exactly
true for the ion case. As a consequence the formalism of the Bargmann spaces
[17], which has been extended from the linear oscillators to the q-ones,
cannot be applied exactly to the ion case. In fact, it will be shown in the
following that this can be done by considering a class of rational
deformations which approximate to any degree of accuracy the ion
deformation. In most experimental cases the statistical state of a trapped
ion is limited to a finite number of Fock states so that these rational
deformations may adequately approximate the ion deformation. Only in this
''weak'' sense it is possible to construct a ion-analogue of a Bargmann
space, on which the deformed creation and annihilation operators are
represented as multiplication by $z$ and differentiation with respect to $z$%
, respectively.

This paper is dedicated to an extension of the theory of the usual coherent
states to NCS using as examples the deformation relative to the dark states
of trapped ions. We start with a single-mode excitation field $A$ (Sec. II),
by discussing some properties of NCS, and introducing a deformed version $%
D_{h}\left( \alpha \right) $ of the displacement operator (Sec. III). In
Sec. IV we discuss some aspects of the resolution of unity for these NCS.
The operator $D_{h}\left( \alpha \right) $ is used in Sec. V for associating
the density matrix operator $\hat{\rho}$ to a linear functional $\rho
_{A,h}\left( z\right) $ mapping the test function $\exp \left( \alpha
z^{*}-\alpha ^{*}z\right) $ into the expectation value $\left\langle
D_{h}\left( \alpha \right) \right\rangle $, by extending the construction of
the antinormal probability distribution function [18]. The connection with
the P-representation is also briefly examined. Section VI is dedicated to
NCS on a circle, for which the Wigner functions are presented. Finally, the
last section is dedicated to the dark states, arising when a trapped ion is
driven by a bichromatic laser field. An asymptotic expression of the
deformation and the relative factorial is obtained and its implication on
the convergence of the NCS series is discussed. It comes out that it
converges only for $\alpha $ in a circle of radius equal to the inverse of $%
\eta $. On the other hand the weight of each Fock state can take values so
large to prevent the resolution of unity in terms of normalized NCS. Some
approximate expressions of the deformation are discussed together with the
possibility of using these NCS for representing the ion statistical state.

\section{\bf Motion of a trapped and laser-driven ion}

We consider an ideal two-level ion of mass M constrained to move in a 3D
harmonic potential. Taking the principal trap (x-axis) axis to coincide with
the direction of propagation of the driving field, one quantum number
suffices to label the vibrational states of the trap. The other two are
traced out by summing over the corresponding degrees of freedom.

The ion's internal and external degrees of freedom are coupled together by a
light field ${\cal E}e^{i\omega _{L}t+i\varphi \left( t\right) }$
periodically modulated at the frequency $\nu $ of the ion trap 
\[
E\left( x,t\right) ={\cal E}e^{i\omega _{L}t+i\varphi \left( t\right)
}g\left( t\right) f\left( x\right) +h.c. 
\]
where $g\left( t\right) =g\left( t+\frac{2\pi }{\nu }\right) $ is a
generally complex periodic function of frequency $\nu $ and $h.c.$ stays for
the Hermitian conjugate. The function $f\left( x\right) $ stands for $%
e^{-ik_{L}x}$ or $\sin \left( k_{L}x+\phi \right) $ respectively for a
progressive or standing wave, with the phase $\phi $ determining the
position of the trap potential with respect to the standing wave.

We will dwell on monochromatic 
\[
g_{N+1}^{\left( 1\right) }\left( t\right) =e^{-i\left( N+1\right) \nu t} 
\]
and bichromatic driving fields 
\[
g_{N+1}^{\left( 2\right) }\left( t\right) =e^{-i\left( N+1\right) \nu
t}-\alpha _{N+1} 
\]
with the parameter $N$ taking non-negative integer values, and $\alpha
_{N+1} $ a complex coefficient depending on the amplitudes of the two waves.

Now, introducing the Lamb-Dicke parameter $\eta =\hbar k_{L}/\sqrt{2M\hbar
\nu }$ we put as usual $e^{-ikx}=e^{-i\eta \left( a_{v}^{\dagger
}+a_{v}\right) }$. In the classical limit $\eta $ is large and the
absorption or emission of a photon will always cause some change in the
vibrational state of the atom. In the non-classical Lamb-Dicke limit (LDL)
of small $\eta $, many photons may need to be absorbed or emitted before the
atom changes vibrational state. For example in the sideband cooling
experiment carried out by Diedrich et al. [19] the parameter $\eta $ was
equal to $0.06$.

The Hamiltonian for a trapped ion interacting with a bichromatic field can
be split in two parts 
\[
H=H_{0}+H_{int} 
\]
where ($\hbar =1$) 
\begin{equation}
H_{0}=\omega _{12}\sigma _{3}+\nu \hat{n}  \label{zero-hamiltonian}
\end{equation}
and, in the electric dipole approximation, 
\[
H_{int}=\wp \left( \sigma _{-}E^{*}\left( x,t\right) +\sigma _{+}E\left(
x,t\right) \right) 
\]

When the Rabi frequency $\Omega $, relative to the laser induced transition
between the ion ground and excited levels, is much smaller than the trapping
potential frequency $\nu $, a perturbation expansion can be carried out in $%
\Omega /\nu $, as discussed in Ref. [9]. This expansion allows a division
into quickly and slowly varying density operator matrix elements, the former
of which can be adiabatically eliminated.

Arresting the calculation to the zeroth-order in $\Omega /\nu $ amounts to
applying the rotating wave approximation. This approach can be easily
pursued by switching to the interaction picture defined by the unitary
operator $U_{rw}=\exp \left[ -i\left( \omega _{L}\sigma _{3}+\nu \hat{n}%
\right) t\right] $ and retaining in the transformed hamiltonian $H^{\prime }$
the time-independent terms together with the slowly varying phase $\varphi
\left( t\right) $ of the laser field, 
\begin{equation}
H^{\prime }=\left( \Delta -\dot{\varphi}\left( t\right) \right) \sigma
_{3}+\Omega \left( \sigma _{-}A+\sigma _{+}A^{\dagger }\right)
\label{effective-hamiltonian}
\end{equation}
with $\Delta =\omega _{12}-\omega _{L}$ the detuning parameter, $\Omega
=e^{-\eta ^{2}/2}\wp {\cal E}$ the vibronic Rabi frequency and 
\begin{equation}
A=e^{-\eta ^{2}/2}\ \overline{g\left( t\right) f\left[ \eta \left( e^{-i\nu
t}a^{\dagger }+e^{i\nu t}a\right) \right] }  \label{A}
\end{equation}
the bar indicating the time average.

Expanding the factor $e^{-i\eta \left( e^{-i\nu t}a^{\dagger }+e^{i\nu
t}a\right) }$ in power series in $a$ and $a^{\dagger }$, introducing the
operator 
\begin{equation}
f_{k}\left( \hat{n},\eta ^{2}\right) =\sum_{m=0}^{\infty }\frac{\left( \hat{n%
}-m+1\right) _{m}}{\left( k+1\right) _{m}m!}\left( -\eta ^{2}\right) ^{m}=k!%
\frac{L_{\hat{n}}^{k}\left( \eta ^{2}\right) }{\left( \hat{n}+1\right) _{k}}
\label{operator-fk}
\end{equation}
with $\left( \hat{n}-m+1\right) _{m}=\left( a^{\dagger }\right) ^{m}a^{m}=%
\hat{n}\left( \hat{n}-1\right) \ldots \left( \hat{n}-m+1\right) $ and $L_{%
\hat{n}}^{k}\left( \eta ^{2}\right) $ reducing in the Fock basis to the
generalized Laguerre polynomials, we obtain respectively for progressive 
\begin{equation}
e^{-i\eta \left( e^{-i\nu t}a^{\dagger }+e^{i\nu t}a\right) }=e^{-\eta
^{2}/2}\sum_{k=0}^{\infty }\epsilon _{k}\frac{\left( -i\eta \right) ^{k}}{k!}%
\left[ f_{k}\left( \hat{n}\right) a^{k}e^{ik\nu t}+\left( a^{\dagger
}\right) ^{k}f_{k}\left( \hat{n}\right) e^{-ik\nu t}\right]
\label{exp-progr}
\end{equation}
and standing waves 
\begin{equation}
\sin \left[ \eta \left( a^{\dagger }+a\right) +\phi \right] =e^{-\eta
^{2}/2}\sum_{k=0}^{\infty }\epsilon _{k}\frac{\left( -\eta \right) ^{k}}{k!}%
\sin \left( \phi +k\frac{\pi }{2}\right) \left[ a^{k}f_{k}\left( \hat{n}%
+k\right) e^{ik\nu t}+f_{k}\left( \hat{n}+k\right) \left( a^{\dagger
}\right) ^{k}e^{-ik\nu t}\right]
\end{equation}
with $k$ a positive integer and $\epsilon _{k}=\frac{1}{2}$ for $k=0$ and $%
\epsilon _{k}=1$ otherwise.

For progressive $\left( p\right) $ and stationary $\left( s\right) $
monochromatic waves with $g\left( t\right) =e^{-i\left( N+1\right) \nu t}$
the operator $A$ (see (\ref{A})) is given by 
\begin{eqnarray}
A_{p}^{\left( 1\right) } &=&\frac{\left( -i\eta \right) ^{N+1}}{\left(
N+1\right) !}f_{N+1}\left( \hat{n},\eta ^{2}\right) a^{N+1}  \nonumber \\
A_{s}^{\left( 1\right) } &=&\left( -i\right) ^{N+1}\sin \left( \phi +(N+1)%
\frac{\pi }{2}\right) A_{p}^{\left( 1\right) }  \label{mono}
\end{eqnarray}
while for two modes bichromatic driving fields

\begin{eqnarray}
A_{p}^{\left( 2\right) } &=&f_{N+1}\left( \hat{n}\right) a^{N+1}-\alpha
_{N+1}f_{0}\left( \hat{n}\right)  \nonumber \\
A_{s}^{\left( 2\right) } &=&\left( -i\right) ^{N+1}\sin \left( \phi +(N+1)%
\frac{\pi }{2}\right) f_{N+1}\left( \hat{n}\right) a^{N+1}+\sin \left( \phi
\right) \alpha _{N+1}f_{0}\left( \hat{n}\right) .  \label{bichrom}
\end{eqnarray}

\section{\bf Nonlinear coherent states }

Coherent states were originally introduced as eigenstates of the
annihilation operator for the harmonic oscillator [20]. They have been
generalized (see [3, 4, 14, 16]) by labeling as nonlinear coherent states $%
\left| \alpha ,h\right\rangle $ the right-hand eigenstates

\begin{equation}
A\left| \alpha ,h\right\rangle =\alpha \left| \alpha ,h\right\rangle
\label{1-coh}
\end{equation}
of operators\footnote{%
for the sake of notational simplicity we will use the same symbol $A$ for
indicating fields of the form (\ref{mono}) and (\ref{bichrom}).} $A$ of the
form 
\begin{equation}
A=ah\left( \hat{n}\right)  \label{non-linear}
\end{equation}
where $h\left( \hat{n}\right) $ is an operator-valued real function of the
number operator. It is immediate to show that 
\begin{equation}
\left| \alpha ,h\right\rangle =N_{h,\alpha }\sum_{n=0}^{\infty }\frac{\alpha
^{n}}{\sqrt{n!}\left[ h\left( n\right) \right] !}\left| n\right\rangle
\label{(2-coh)}
\end{equation}
with $\left[ h\left( n\right) \right] !=h\left( 0\right) h\left( 1\right)
\cdots h\left( n\right) $ and normalizing factor $N_{h,\alpha }$ 
\[
N_{\alpha ,h}=\frac{1}{\sqrt{E_{h}\left( \left| \alpha \right| ^{2}\right) }}
\]
expressed in terms of the entire function 
\[
E_{h}\left( v\right) =\sum_{n=0}^{\infty }\frac{v^{n}}{n!\left( \left[
h\left( n\right) \right] !\right) ^{2}} 
\]
referred to in the following as h-exponential in analogy with the
q-exponential used in Ref. [3].

The deformation functions $h\left( n,\eta ^{2}\right) $ associated to the
dark states of the trapped ions are represented by the ratio of two Laguerre
polynomials of argument equal to the Lamb--Dicke parameter $\eta ^{2}$ so
that they vanish or become infinite for some isolated combinations of $\eta
^{2}$ and $n$. We are obliged to explicitly assume that this situation does
not occur for the values of $\eta ^{2}$ considered.

In Sec. VII we will obtain an asymptotic expression of the weights of the
Fock states occurring in the series expansion of the NCS relative to trapped
ions. They take very large and very small values for increasing $n$, so that
these NCS can be normalized only for $\alpha $ inside the circle $1/\eta .$
For convenience of discussion we shall ignore this problem by restricting
our treatment here to normalized NCS states.

It may be worth noting at this point that many of the foregoing formulas may
be abbreviated by adopting a normalization different from the conventional
one for the coherent state. If we introduce the symbol $\left\| \alpha
;h\right\rangle $ for the states normalized in the new way and define these
as 
\[
\left\| \alpha ;h\right\rangle =N_{\alpha ,h}^{-1}\left| \alpha
;h\right\rangle 
\]
and

\[
\left\langle \alpha ;h\right. \left\| \beta ;h\right\rangle =E_{h}\left(
\alpha ^{*}\beta \right) . 
\]

Since the commutator $\left[ A^{\dagger },A\right] =\hat{n}h^{2}\left( \hat{n%
}\right) -(\hat{n}+1)h^{2}(\hat{n}+1)$ is not a c-number it is worthy
introducing the operator [16] 
\begin{equation}
A_{h}^{\dagger }=\frac{1}{h\left( \hat{n}\right) }a^{\dagger }
\label{A-deformed}
\end{equation}

With these alterations we have 
\begin{eqnarray*}
A\left\| \alpha ,h\right\rangle &=&\alpha \left\| \alpha ,h\right\rangle , \\
\hat{n}\left\| \alpha ,h\right\rangle &=&\alpha \frac{\partial }{\partial
\alpha }\left\| \alpha ,h\right\rangle , \\
A_{h}^{\dagger }\left\| \alpha ,h\right\rangle &=&\frac{\partial }{\partial
\alpha }\left\| \alpha ,h\right\rangle .
\end{eqnarray*}
In addition 
\begin{equation}
A^{\dag }\left\| \alpha ,h\right\rangle =h^{2}\left( \hat{n}\right)
A_{h}^{\dag }\left\| \alpha ,h\right\rangle =h^{2}\left( \alpha \frac{%
\partial }{\partial \alpha }\right) \frac{\partial }{\partial \alpha }%
\left\| \alpha ,h\right\rangle  \label{adjoint-coherent}
\end{equation}
In all the above r.h.s. the operators $\alpha ,\partial _{\alpha }$ and
their combination are intended to act on the coefficients of the Fock states
series.

\subsection{\bf Displacement and deformation operators}

It is well known that coherent state $\left| \alpha \right\rangle $ can be
also introduced by displacing the Fock vacuum state $\left| 0\right\rangle $
by means of the operator 
\begin{equation}
{\cal D}\left( \alpha \right) =\exp \left( -\alpha ^{*}a+\alpha a^{\dagger
}\right)  \label{displ-linear}
\end{equation}
due to its property of displacing the annihilation operator $a$ by the
generally complex quantity $\alpha $, 
\[
{\cal D}\left( \alpha \right) a{\cal D}\left( -\alpha \right) =a-\alpha 
\]
Unfortunately, ${\cal D}\left( \alpha \right) $ is unable to displace the
deformed operator $A$. In alternative ${\cal D}\left( \alpha \right) $ could
be replaced by the unitary operator obtained by replacing in Eq. (\ref
{displ-linear}) $a$ and $a^{\dagger }$ by $A$ and $A^{\dagger }$
respectively, but also this operator does not displace $A$ by the complex
quantity $\alpha $. The difficulties in dealing with exponentials of linear
combinations of $A$ and $A^{\dagger }$ originate from the circumstance that
their commutator is not a c-number. These problems can be overcome by using $%
A_{h}^{\dagger }$ (see Eq. (\ref{A-deformed})) in place of $A^{\dagger }$
and defining the ''deformed'' version of the displacement operator as 
\begin{equation}
{\cal D}_{h}\left( \alpha \right) =\exp \left( -\alpha ^{*}A+\alpha
A_{h}^{\dagger }\right) =e^{\frac{\left| \alpha \right| ^{2}}{2}}e^{-\alpha
^{*}A}e^{\alpha A_{h}^{\dagger }}=e^{-\frac{\left| \alpha \right| ^{2}}{2}%
}e^{\alpha A_{h}^{\dagger }}e^{-\alpha ^{*}A}  \label{displ-deform}
\end{equation}
${\cal D}_{h}\left( \alpha \right) $ shares many properties of the standard
operator ${\cal D}\left( \alpha \right) $ as 
\[
{\cal D}_{h}^{-1}\left( \alpha \right) ={\cal D}_{h}\left( -\alpha \right) 
\]
and 
\[
{\cal D}_{h}\left( \beta \right) {\cal D}_{h}\left( \alpha \right) =\exp
\left[ \frac{1}{2}\left( \beta \alpha ^{*}-\beta ^{*}\alpha \right) \right] 
{\cal D}_{h}\left( \beta +\alpha \right) 
\]
However, ${\cal D}_{h}\left( \alpha \right) $ is not a unitary operator, 
\[
{\cal D}_{h}^{\dagger }\left( \alpha \right) ={\cal D}_{\frac{1}{h}%
}^{-1}\left( \alpha \right) ={\cal D}_{\frac{1}{h}}\left( -\alpha \right) 
\]
so that it does not preserve the norm of a state.

${\cal D}_{h}\left( \alpha \right) $ and ${\cal D}_{\frac{1}{h}}\left(
\alpha \right) $ displace $A$ and $A^{\dagger }$ respectively by $\alpha $
and $\alpha ^{*}$, 
\begin{eqnarray}
{\cal D}_{h}\left( \alpha \right) A{\cal D}_{h}\left( -\alpha \right)
&=&A-\alpha  \nonumber \\
{\cal D}_{\frac{1}{h}}\left( \alpha \right) A^{\dagger }{\cal D}_{\frac{1}{h}%
}\left( -\alpha \right) &=&A^{\dagger }-\alpha ^{*}  \label{displaced-1}
\end{eqnarray}
and $A_{h}^{\dagger }$ by $\alpha ^{*}$%
\begin{equation}
{\cal D}_{h}\left( \alpha \right) A_{h}^{\dagger }{\cal D}_{h}\left( -\alpha
\right) =A_{h}^{\dagger }-\alpha ^{*}  \label{displaced-2}
\end{equation}

Accordingly, the NCS $\left| \alpha ;h\right\rangle $ can be obtained by
applying ${\cal D}_{h}\left( \alpha \right) $ to the vacuum state, 
\[
\left\| \alpha ;h\right\rangle =e^{\alpha A_{h}^{\dagger }}\left|
0\right\rangle =e^{\frac{\left| \alpha \right| ^{2}}{2}}{\cal D}_{h}\left(
\alpha \right) \left| 0\right\rangle 
\]

In conclusion, the NCS $\left\| \alpha ;h\right\rangle $ can be obtained by
deforming the usual coherent state $\left\| \alpha \right\rangle $ by means
of the deformation operator 
\begin{equation}
d_{h}={\cal D}_{h}\left( \alpha \right) {\cal D}\left( -\alpha \right)
\label{def-operator}
\end{equation}
namely, 
\[
\left\| \alpha ;h\right\rangle =d_{h}\left\| \alpha \right\rangle 
\]
Although expressed as a product of operators depending on the complex
parameter $\alpha ,$ $d_{h}$ is independent of $\alpha $. In a Fock basis it
is diagonal with components equal to $\left[ h\left( n\right) \right] !^{-1}$%
. Since $h\left( n\right) $ does not vanish, as already assumed, $d_{h}$ is
not singular.

Finally, we note that 
\[
\left\langle m\left| {\cal D}_{h}\left( \alpha \right) \right|
n\right\rangle =\frac{\left[ h\left( m\right) \right] !}{\left[ h\left(
n\right) \right] !}\left\langle m\left| {\cal D}\left( \alpha \right)
\right| n\right\rangle 
\]
so that the matrix representation of ${\cal D}$ and ${\cal D}_{h}$ have the
same diagonal part.

A further remark is that the set of operators ${\cal D}_{h}(\alpha )$
constitutes a Weyl system which does not lead to the canonical quantization
for not being unitary.

\subsection{\bf Nonlinear displaced Fock states}

In Sec. V we will use the Fock states displaced by ${\cal D}_{h}\left(
\alpha \right) $ (see Eq. (\ref{lin-func})) 
\begin{equation}
\left| \varphi _{m},\alpha ,h\right\rangle ={\cal D}_{h}\left( \alpha
\right) \left| m\right\rangle  \label{displ-states}
\end{equation}
which can be shown with the help of Eqs. (\ref{displaced-1}) and (\ref
{displaced-2}) to be the right eigenstates of the operator $\left(
A_{h}^{\dagger }-\alpha ^{*}\right) \left( A-\alpha \right) ={\cal D}%
_{h}\left( \alpha \right) \hat{n}{\cal D}_{h}\left( -\alpha \right) $, 
\[
\left( A_{h}^{\dagger }-\alpha ^{*}\right) \left( A-\alpha \right) \left|
\varphi _{m},\alpha ,h\right\rangle =m\left| \varphi _{m},\alpha
,h\right\rangle 
\]
Analogously we can introduce the left eigenstates defined by 
\[
\left\langle \psi _{m},\alpha ,h\right| \left( A_{h}^{\dagger }-\alpha
^{*}\right) \left( A-\alpha \right) =m\left\langle \psi _{m},\alpha
,h\right| 
\]
which are obtained by displacing $\left\langle m\right| $ by ${\cal D}%
_{h}\left( -\alpha \right) ,$ i.e. $\left\langle \psi _{m},\alpha ,h\right|
=\left\langle m\right| {\cal D}_{h}\left( -\alpha \right) .$

It is noteworthy that the left and right displaced Fock states are mutually
orthogonal, 
\[
\left\langle \psi _{m},\alpha ,h\right| \left. \varphi _{n},\alpha
,h\right\rangle =0 
\]
for $m\neq n.$

On the other hand these states can be also expressed in the form 
\begin{eqnarray*}
\left| \varphi _{m},\alpha ,h\right\rangle &=&\frac{\left[ h\left( m\right)
\right] !\left( A_{h}^{\dagger }-\alpha ^{*}\right) ^{m}}{\sqrt{m!}}\left|
\alpha ,h\right\rangle =\frac{\left[ h\left( m\right) \right] !}{\sqrt{m!}}%
\sum_{n}%
{m \choose n}%
\left( -\alpha ^{*}\right) ^{m-n}\left| \alpha ,h,m\right\rangle \\
\left\langle \psi _{m},\alpha ,h\right| &=&\left\langle m\right| D_{h}\left(
-\alpha \right) =\left\langle \alpha ,h\right| \frac{\left( A-\alpha \right)
^{m}}{\left[ h\left( m\right) \right] !\sqrt{m!}}=\frac{1}{\left[ h\left(
m\right) \right] !\sqrt{m!}}\sum_{n}%
{m \choose n}%
\left( -\alpha \right) ^{m-n}\left\langle \alpha ,h,m\right|
\end{eqnarray*}
where $\left| \alpha ,h,m\right\rangle =A_{h}^{\dagger m}\left| \alpha
,h\right\rangle $ and $\left\langle \alpha ,h,m\right| =\left\langle \alpha
,h\right| A^{m}$ stand for the deformed versions of the excited coherent
states [20] (see also [21]).

\section{\bf Resolution of the unity}

From the completeness relation of coherent states 
\[
1=\frac{1}{\pi }\int \left| \alpha \right\rangle \left\langle \alpha \right|
d^{2}\alpha ~ 
\]
it descends 
\[
1=\frac{1}{\pi }\int ~d_{h}\left| \alpha \right\rangle \left\langle \alpha
\right| d_{h}^{-1}d^{2}\alpha =\frac{1}{\pi }\int ~d_{h}^{-1}\left| \alpha
\right\rangle \left\langle \alpha \right| d_{h}d^{2}\alpha 
\]
Next, using the relation 
\[
d_{h}^{-1}={\cal D}\left( \alpha \right) {\cal D}_{\frac{1}{h}}^{\dagger
}\left( \alpha \right) =\left( {\cal D}_{\frac{1}{h}}\left( \alpha \right) 
{\cal D}\left( -\alpha \right) \right) ^{\dagger }=d_{\frac{1}{h}}^{\dagger
} 
\]
the above resolution of unity can be expressed in terms of deformed coherent
states 
\begin{equation}
1=\frac{1}{\pi }\int \frac{e^{-\alpha \alpha *}}{N_{\alpha ,h}N_{\alpha ,%
\frac{1}{h}}}\left| \alpha ,h\right\rangle \left\langle \alpha
,h^{-1}\right| d^{2}\alpha ~=\frac{1}{\pi }\int ~\frac{e^{-\alpha \alpha *}}{%
N_{\alpha ,h}N_{\alpha ,\frac{1}{h}}}\left| \alpha ,h^{-1}\right\rangle
\left\langle \alpha ,h\right| d^{2}\alpha  \label{asymmetric-resolution}
\end{equation}
It goes without saying that this resolution holds true only if the NCS
relative to the deformations $h$ and $1/h$ are both normalizable in the
whole complex $\alpha $--plane.

For some deformations anyhow it is possible to obtain a resolution of unity
in terms of projectors of deformed coherent states, i.e. to find a suitable
element of measure $d\mu $ such that 
\begin{equation}
1=\int \left\| \alpha ,h\right\rangle \left\langle \alpha ,h\right\| d\mu
\label{unity-resolution}
\end{equation}
$d\mu $ can be considered as an extension of the measure element $d\mu =%
\frac{1}{\pi }e^{-\left| \alpha \right| ^{2}}d^{2}\alpha $ [17] for the
linear oscillators. Since $\int \left\langle m\right. \left\| \alpha
,h\right\rangle \left\langle \alpha ,h\right\| \left. n\right\rangle d\mu $
must vanish for $m\neq n$ $d\mu $ can be put in the form 
\[
d\mu =\frac{1}{\pi }m_{h}\left( \left| \alpha \right| ^{2}\right)
d^{2}\alpha 
\]
where $m_{h}\left( x\right) $ is a distribution satisfying the set of
equations 
\begin{equation}
n!\left( \left[ h\left( n\right) \right] !\right) ^{2}=\int m_{h}\left(
x\right) x^{n}dx  \label{integral-equation-mu}
\end{equation}
for every integer $n$.

Treating $n=s-1$ as a continuous variable the above relation represents a
Mellin integral transform, 
\begin{equation}
g\left( s\right) =\int_{0}^{\infty }f\left( x\right) x^{s-1}dx
\label{Mellin-transform-1}
\end{equation}
so that $m_{h}\left( x\right) $ is the Mellin antitransform of $g\left(
s\right) =\Gamma \left( s\right) \left( \left[ h\left( s-1\right) \right]
!\right) ^{2}$.

From the relation $\left\langle \beta ,h\right| A^{m}\left| \beta
,h\right\rangle =\beta ^{m}$ it descends that $E_{h}\left( \beta ^{*}\alpha
\right) $ is the self-reproducing kernel of the h-analogue of the Bargmann
space [17], with respect to $d\mu $%
\[
\int \left| E_{h}\left( \beta ^{*}\alpha \right) \right| ^{2}\alpha ^{m}d\mu
=\beta ^{m} 
\]

In preparation of the discussion of Sec. VII it is worth remarking that
replacing $h$ by the deformation $\beta h$ the relative measure $m_{\beta
h}\left( x\right) $ is given by 
\begin{equation}
m_{\beta h}\left( x\right) =\beta ^{-2}m_{h}\left( \beta ^{-2}x\right)
\label{measure-scaling}
\end{equation}
This relation can be also used for expressing a thermal density matrix
characterized by Boltzmann weight factors $\rho _{nn}\varpropto \exp \left(
-\beta n\right) $ in the form 
\[
\hat{\rho}=e^{2\beta }\int \frac{m_{h}\left( e^{2\beta }\left| \alpha
\right| ^{2}\right) }{m_{h}\left( \left| \alpha \right| ^{2}\right) }\left\|
\alpha ,h\right\rangle \left\langle \alpha ,h\right\| d\mu 
\]

In Ref. [3] it was possible to obtain the resolution of unity for a
q-oscillator by deforming both the derivative and the integral operators
while a resolution for the so-called harmonious states was obtained in [23].
We will see in the following that for the trapped ion deformation the
measure is a distributional Laplace antitransform which includes
non-normalizable NCS.

For a deformation approximated by a rational function of $n$, $g\left(
s\right) $ corresponds to the ratio of products of gamma functions, 
\begin{equation}
g\left( s\right) =\frac{\Gamma \left( a_{1}+s\right) \cdots \Gamma \left(
a_{A}+s\right) }{\Gamma \left( b_{1}+s\right) \cdots \Gamma \left(
b_{B}+s\right) }\equiv \Gamma \left[ 
{\left( a\right) +s \atop \left( b\right) +s}%
\right]  \label{fact-gen}
\end{equation}
For $A\geq B$ the relative antitransform is given by a combination of
generalized hypergeometric functions 
\[
_{C}F_{D}\left[ 
{\left( c\right)  \atop \left( d\right) }%
;\left( -1\right) ^{C+D+1}x\right] =\sum_{n}\frac{\left( c_{1}\right)
_{n}\cdots \left( c_{C}\right) _{n}}{\left( d_{1}\right) _{n}\cdots \left(
d_{D}\right) _{n}}\frac{x^{n}}{n!}\left( -1\right) ^{n\left( C+D+1\right) } 
\]
namely [24] 
\begin{equation}
m_{h}\left( x\right) =\sum_{\mu =1}^{A}\Gamma \left[ 
{\left( a\right) ^{\prime }-a_{\mu } \atop \left( b\right) -a_{\mu }}%
\right] \ _{B}F_{A-1}\left[ 
{1+a_{\mu }-\left( b\right)  \atop 1+a_{\mu }-\left( a\right) ^{\prime }}%
;\left( -1\right) ^{A+B}x\right] x^{a_{\mu }}  \label{Mellin-anti}
\end{equation}
where $\left( a\right) ^{\prime }-a_{\mu }$ and $\left( a\right) ^{\prime
}-\left( a\right) ^{\prime }-a_{\mu }-1$ stand for the sequences $%
a_{1}-a_{\mu },\ldots ,a_{A}-a_{\mu },$ and $a_{1}-a_{\mu }-1,\ldots
,a_{A}-a_{\mu }-1$ with the exclusion the $\mu $-th term.

\section{\bf Expansion of statistical states}

The same reasons that led [22] to express arbitrary states and operators in
term of coherent states, suggest that we develop expansions in terms of NCS
as well.

Following [18] we introduce for a statistical state the deformed quantum
linear functional 
\begin{equation}
F_{h}\left[ \alpha \right] =Tr\left\{ \hat{\rho}{\cal D}_{h}\left( \alpha
\right) \right\} =\sum_{m}\rho _{mn}\left\langle n\right| \left. \varphi
_{m},\alpha ,h\right\rangle  \label{lin-func}
\end{equation}
In particular for a diagonal density matrix $F_{h}\left[ \alpha \right] $
reduces to the standard $F\left[ \alpha \right] .$

Using the unity resolution (\ref{asymmetric-resolution}) $F_{h}\left[ \alpha
\right] $ may be rewritten as 
\begin{equation}
F_{h}\left[ \alpha \right] =e^{\frac{\alpha \alpha *}{2}}\int \exp \left(
\alpha z^{*}-\alpha ^{*}z\right) \rho _{h,A}\left( z\right) d^{2}z
\label{lin-func-bis}
\end{equation}
where 
\begin{eqnarray}
\rho _{h,A}\left( z\right) &=&\frac{1}{\pi }\frac{e^{-zz^{*}}}{%
N_{z,h}N_{z,h^{-1}}}Tr\left\{ \hat{\rho}\left| z,h\right\rangle \left\langle
z,h^{-1}\right| \right\}  \nonumber \\
&=&\frac{1}{\pi }\left\langle z\left| \hat{\rho}_{h}\right| z\right\rangle
\label{antin-matrix}
\end{eqnarray}
with 
\[
\hat{\rho}_{h}=d_{h}^{-1}\hat{\rho}d_{h} 
\]
the deformed density operator. In other words $\rho _{h,A}\left( z\right) $
stands for the generalized distribution function of the deformed density
matrix $d_{h}^{-1}\hat{\rho}d_{h}$.

For extending the definition of the characteristic functional $F\left[
\alpha \right] =Tr\left\{ \hat{\rho}D\left( \alpha \right) \right\} $ to a
deformed oscillator we pay the penalty of loosing some properties of $\rho
_{A}\left( z\right) $. In fact, $\rho _{h,A}\left( z\right) $ may take in
general positive and negative values. It can be regarded as a generalized
probability distribution function as long as the association between
operators and functions is based on antinormal ordering, 
\begin{equation}
Tr\left\{ \hat{\rho}G_{A}\left( A,A_{h}^{\dagger }\right) \right\} =\int
\rho _{h,A}\left( z\right) G_{A}\left( z,z^{*}\right) d^{2}z
\label{antinormal-expectation}
\end{equation}

In particular, for a diagonal density matrix $\rho _{h,A}\left( z\right)
=\rho _{A}\left( z\right) $ while for $\hat{\rho}=\left| w,h\right\rangle
\left\langle w,h\right| $ 
\[
\rho _{h,A}\left( z\right) =\frac{1}{\pi }\frac{E_{h}\left( w^{*}z\right) }{%
E_{h}\left( w^{*}w\right) }\exp \left[ z^{*}\left( w-z\right) \right] 
\]
Consequently, the transformation (\ref{lin-func-bis}) applies if there
exists the Fourier transform of $\exp \left[ z^{*}\left( w-z\right) \right]
E_{h}\left( w^{*}z\right) $. Analogously, working with $\hat{\rho}=\left|
w,h^{-1}\right\rangle \left\langle w,h^{-1}\right| $ we arrive at the same
conclusion for $\exp \left[ z^{*}\left( w-z\right) \right] E_{h^{-1}}\left(
w^{*}z\right) $. This in turn implies that $E_{h^{-1}}\left( w^{*}z\right) $
and $E_{h}\left( w^{*}z\right) $ cannot grow at infinity as quickly as $\exp
\left( zz^{*}\right) $.

We recall that in the coherent states representation of a bounded operator $%
\hat{O}$, the vanishing of $\left\langle z\left| \hat{O}\right|
z\right\rangle =0$ in a domain of the complex plane of finite area implies
the vanishing of $\hat{O}$ itself (see Refs. [18]). Since $d_{h}$ has been
assumed non singular the same theorem holds true for $\left\langle z\left| 
\hat{O}_{h}\right| z\right\rangle $, so that two deformed density matrices
having the same function $\rho _{h,A}\left( z\right) $ over some area of $z$%
, must coincide. In conclusion, Eq. (\ref{antin-matrix}) establishes a
one-to-one correspondence between the operator $\hat{\rho}$ and the function 
$\rho _{h,A}\left( z\right) $.

When the deformation admits the unity resolution (\ref{unity-resolution}) a
density matrix can be represented in several cases by a P-representation, 
\begin{equation}
\hat{\rho}=\int P_{h}\left( \alpha \right) \left\| \alpha ,h\right\rangle
\left\langle \alpha ,h\right\| d\mu  \label{P-representation}
\end{equation}
in which $P_{h}\left( \alpha \right) $ can be regarded as a generalized
probability distribution function as long as the association between
operators and functions is based on normal ordering, 
\begin{equation}
Tr\left\{ \hat{\rho}G_{N}\left( A^{\dagger },A\right) \right\} =\int
P_{h}\left( \alpha \right) G_{N}\left( \alpha ,\alpha ^{*}\right) d\mu
\end{equation}

When $\hat{\rho}$ is represented in the form (\ref{P-representation}) the
master equation of $\hat{\rho}$ can be in many cases transformed in a master
equation for $P_{h}.$ This circumstance becomes particularly valuable in the
study of the decay of an excited trapped ion toward the fundamental dark
state. In this case we are faced for example with operators of the form 
\begin{eqnarray*}
A^{\dag }\hat{\rho} &=&\int P\left( \alpha \right) m_{h}\left( \alpha \alpha
^{\star }\right) h^{2}\left( \alpha \frac{\partial }{\partial \alpha }%
\right) \frac{\partial }{\partial \alpha }\left\| \alpha \right\rangle
\left\langle \alpha \right\| d^{2}\alpha \\
&=&-\int \left\| \alpha \right\rangle \left\langle \alpha \right\| \frac{%
\partial }{\partial \alpha }h^{2}\left( -1-\alpha \frac{\partial }{\partial
\alpha }\right) \left\{ P\left( \alpha \right) m_{h}\left( \alpha \alpha
^{\star }\right) \right\} d^{2}\alpha
\end{eqnarray*}
use having been made of Eq. $\left( \ref{adjoint-coherent}\right) .$
Expanding $P\left( \alpha \right) m_{h}\left( \alpha \alpha ^{\star }\right) 
$ in power series of $\alpha ^{m}\left( \alpha ^{\star }\right) ^{n}$ we see
that 
\[
\frac{\partial }{\partial \alpha }h^{2}\left( -1-\alpha \frac{\partial }{%
\partial \alpha }\right) \alpha ^{m}\left( \alpha ^{\star }\right)
^{n}=mh^{2}\left( -1-m\right) \alpha ^{m-1}\left( \alpha ^{\star }\right)
^{n} 
\]

\section{\bf Nonlinear coherent states on a circle}

The above definition of NCS states (we will call them of order 1) can be
extended to the eigenstates of the operators $A_{N+1}$ of a more general
form $A_{N+1}=a^{N+1}h\left( \hat{n}\right) $ [25] and so the equation 
\[
A_{N+1}\left| \alpha ,h,N+1,q\right\rangle =\alpha \left| \alpha
,h,N+1,q\right\rangle 
\]
with $N>0$ is considered.

The eigenstate belonging to the eigenvalue $\alpha $ is $N+1$--fold
degenerate and $q$ is an integer ranging from 0 to $N$. In terms of Fock
states we have 
\begin{equation}
\alpha \left| \alpha ,h,N+1,q\right\rangle =N_{\alpha
,h,q}\sum_{l=0}^{\infty }\frac{\alpha ^{l\left( N+1\right) +q}}{\sqrt{\left(
l\left( N+1\right) +q\right) !}\left[ h\left( l\left( N+1\right) +q\right)
\right] !}\left| l\left( N+1\right) +q\right\rangle  \label{N+1-coh}
\end{equation}
with the normalization factor 
\[
\left| N_{\alpha ,h,q}\right| ^{-2}=\sum_{l=0}^{\infty }\frac{\left| \alpha
\right| ^{2l\left( N+1\right) +2q}}{\left( l\left( N+1\right) +q\right)
!\left( \left[ h\left( l\left( N+1\right) +q\right) \right] !\right) ^{2}} 
\]
where 
\[
\left[ h\left( l\left( N+1\right) +q\right) \right] !=h\left( q\right)
h\left( N+1+q\right) \cdots h\left( l\left( N+1\right) +q\right) . 
\]

Such a state can also be expressed as a sum of NCS (see Eq.\ref{(2-coh)}).
In facts, by introducing the function $h^{\left( N+1\right) }\left( n\right) 
$ defined recursively by 
\[
h^{\left( N+1\right) }\left( l\left( N+1\right) +q\right) =\frac{h\left(
q-1\right) \left[ h\left( l\left( N+1\right) +q\right) \right] !}{h\left(
q\right) \left[ h\left( l\left( N+1\right) +q-1\right) \right] !}h^{\left(
N+1\right) }\left( q\right) 
\]
we have also 
\begin{equation}
\left| \alpha ,h,N+1,q\right\rangle =N_{\alpha ,h,q}^{\prime
}\sum_{k=0}^{N}\left( \epsilon ^{*}\right) ^{qk}\left| \alpha \epsilon
^{k},h^{\left( N+1\right) }\right\rangle  \label{coh-circle}
\end{equation}
with $\epsilon =\exp \left( \frac{i2\pi }{N+1}\right) $ and $N_{\alpha
,h,q}^{\prime }$ a normalization coefficient$.$

We have obtained that a NCS coherent state of order $N+1$ is decomposed in
the sum of $N+1$ first order NCS of complex amplitudes $\alpha ,$ $\alpha
\epsilon ^{q},\ldots ,\alpha \left( \epsilon ^{q}\right) ^{N}$ distributed
uniformly on a circle. These states, referred to as ''crystallized cats'' in
Ref. [26], were introduced for the linear oscillator [27] in the attempt to
generalize the optical Schr\"{o}dinger cats of harmonic oscillators.

Using the deformed displacement operator we have also 
\[
\left| \alpha ,h,N+1,q\right\rangle =N_{\alpha ,h,q}^{\prime }\left(
\sum_{k=0}^{N}\left( \epsilon ^{*}\right) ^{qk}{\cal D}_{h^{\left(
N+1\right) }}\left( \alpha \epsilon ^{k}\right) \right) \left|
0\right\rangle 
\]

In conclusion, the Hilbert space is the direct sum of $N+1$ spaces ${\cal H}=%
{\cal H}_{0}\oplus {\cal H}_{1}\oplus \cdots \oplus {\cal H}_{N}$, $\left(
q=0,1,\ldots N\right) $, each of them having for basis the Fock states $%
\left| \ell \left( N+1\right) +q\right\rangle $, as $l\in (0,...\infty ).$

For $N=1$ the fundamental states of ${\cal H}_{0}$ and ${\cal H}_{1}$ are
respectively the even and odd Schr\"{o}dinger cats. It will be shown in a
following paper that when the radiative damping is negligible, the initial
density matrix separate in the product of two matrices evolving respectively
toward the even and odd Schr\"{o}dinger cats.

\subsection{\bf Wigner function}

The Wigner function ~[27] relative to these states on a circle can be shown
to be given for a generic integer $N$ by 
\begin{eqnarray}
&&W_{N+1}\left( \tilde{q},\tilde{p}\right) =N_{\alpha
,h_{N+1},q}^{2}e^{-\left| q+ip\right| ^{2}}\sum_{ll^{\prime }}\left( \sqrt{2}%
\left( \tilde{q}-i\tilde{p}\right) \right) ^{\left( l^{\prime }-l\right)
\left( N+1\right) }  \label{Wigner} \\
&\times &\frac{\left( -\alpha \right) ^{l\left( N+1\right) +q}}{\left[
h_{N+1}\left( l\left( N+1\right) +q\right) \right] !}\frac{\left( \alpha
^{*}\right) ^{l^{\prime }\left( N+1\right) +q}}{\left[ h_{N+1}\left(
l^{\prime }\left( N+1\right) +q\right) \right] !}\frac{L_{l\left( N+1\right)
+q}^{\left( l^{\prime }-l\right) \left( N+1\right) }\left( 2\left( \tilde{q}%
^{2}+\tilde{p}^{2}\right) \right) }{\left[ l^{\prime }\left( N+1\right)
+q\right] !}  \nonumber
\end{eqnarray}

Analogously for the Husimi--Kano [27] Q-function $Q_{N+1,q}(\tilde{q},\tilde{%
p})=\left| \left\langle \left. \frac{\tilde{q}+i\tilde{p}}{\sqrt{2}}\right|
\alpha ,q,h_{N+1}\right\rangle \right| ^{2}$ 
\begin{equation}
Q_{N+1,q}(\tilde{q},\tilde{p})=N_{\alpha ,h_{N+1},q}^{2}e^{-\frac{\tilde{q}%
^{2}+\tilde{p}^{2}}{2}}\left| \sum_{l=0}^{\infty }\frac{\left( \alpha \frac{%
\tilde{q}-i\tilde{p}}{\sqrt{2}}\right) ^{l\left( N+1\right) +q}}{\left[
l\left( N+1\right) +q\right] !\left[ h_{N+1}\left( l\left( N+1\right)
+q\right) \right] !}\right| ^{2}.  \label{Husimi-Kano}
\end{equation}
with $\left| \frac{\tilde{q}+i\tilde{p}}{\sqrt{2}}\right\rangle $ a
coherent-state vector.

For $N=1$ these states reduce to even $\left( q=0\right) $ and odd $\left(
q=1\right) $ Schr\"{o}dinger cats [15]. In Ref. [28] is examined the
squeezing and antibunching effects by using the function $h_{1}(n)$
introduced in [14] for the NCS. We will see in the following (see Eq. (\ref
{nonlinear-ds})) that the nonlinear cats representing the dark state of a
trapped ion are properly described by the deformation $h_{2}\left( n\right)
=L_{n-2}^{2}\left( \eta ^{2}\right) /[n\left( n-1\right) L_{n-2}\left( \eta
^{2}\right) ]$.

In Fig. 1 we show the Wigner functions for nonlinear even Schr\"{o}dinger
cats of amplitude $\alpha =3.5$ (real) and different parameters $\eta $. In
the linear case ($\eta =0$ Fig. 1-a) the quantum interference is localized
around the origin. The two coherent gaussian peaks are circularly shaped.

For increasing $\eta $ the nonlinearity flattens the interference pattern
while the central interference fringes, particularly their negative part,
become more evident. This is essentially due to a reshaping of the coherent
contribution peak from a gaussian-like nearly circular shape to an
elliptical one with minor axis parallel to the direction connecting the two
coherent peaks. These come closer to the origin and the region where the
Wigner function is non-zero shrinks notably.

A further increase in $\eta $ causes the progressive coming closer and
closer of the main peaks, while the interference fringes become more
localized. For higher $\eta $ there are some interference fringes spreading
over the two coherent peaks. This phenomenon is dominant for very high $\eta 
$ values (Fig. 1 b,c,d $\eta =0.5$) where the main peaks come into the
interference region and the coherent character of the two states forming the
cat is no more distinguishable. The interference area become larger than in
the linear case and a circular symmetry of the interference pattern become
evident.

In Fig. 2 we present two NCS on a circle formed by the superposition of 3
and 4 NCS ($\alpha =3.5$ and $\eta =0.33$).

\section{\bf Dark states}

Cirac et al [13] first proposed in 1993 a scheme for preparing coherent
squeezed states of motion in an ion trap based on the multichromatic
excitation of a trapped ion. Using two waves with beat frequency equal to
twice the trap frequency a ''dark resonance'' appears in the fluorescence
emitted by the ion, the ion is placed in a squeezed state. Similar ''dark
states'' produced by a bichromatic field with beat frequency equal to the
trap frequency were studied by Vogel et al. in 1996 [14] and identified as
nonlinear coherent states.

We will consider in the following a beat frequency which is a generic
multiple of the trap frequency $\nu $, and the ion dark state is described
by a generalized coherent state on a circle. We will consider a bichromatic
field of the type

\begin{equation}
A_{b,N+1}=f_{N+1}\left( \hat{n}\right) a^{N+1}-\alpha _{N+1}f_{0}\left( \hat{%
n}\right) .  \label{two-modes}
\end{equation}
for which the dark state satisfies the equation

\[
\left( a^{N+1}\frac{f_{N+1}\left( \hat{n}-N-1\right) }{f_{0}\left( \hat{n}%
-N-1\right) }-\alpha _{N+1}\right) \left| \psi _{dark}\right\rangle =0. 
\]
that is (see Eq. (\ref{N+1-coh})) 
\[
\left| \psi _{dark}\right\rangle =\left| \alpha
_{N+1},h_{N+1},N+1,q\right\rangle 
\]
with 
\begin{equation}
h_{N+1}\left( \hat{n};\eta ^{2}\right) =\left( N+1\right) !\frac{L_{\hat{n}%
-N-1}^{N+1}}{\left( \hat{n}-N\right) _{N+1}L_{\hat{n}-N-1}}  \label{hN+1}
\end{equation}
and 
\[
\alpha _{N+1}=\frac{\Omega _{0}}{\Omega _{N+1}}\frac{\left( N+1\right) !}{%
\left( -i\eta \right) ^{N+1}} 
\]

In short the dark state is the superposition of $N+1$ nonlinear coherent
states which are equidistantly separated from each other along a circle with
modulation factor $\epsilon ^{k}=\exp \left( \frac{2\pi ik}{N+1}\right) $.

In particular 
\begin{eqnarray}
h_{1}\left( \hat{n};\eta ^{2}\right) &=&\frac{L_{\hat{n}-1}^{1}}{\hat{n}L_{%
\hat{n}-1}}=\frac{L_{\hat{n}-1}-L_{\hat{n}}}{\eta ^{2}L_{\hat{n}-1}} 
\nonumber \\
h_{2}\left( \hat{n};\eta ^{2}\right) &=&\frac{2L_{\hat{n}-2}^{2}}{\hat{n}%
\left( \hat{n}-1\right) L_{\hat{n}-2}}  \label{deformations}
\end{eqnarray}
and 
\[
\alpha _{1}=i\frac{\Omega _{0}}{\Omega _{1}}\;,\;\alpha _{2}=-\frac{\Omega
_{0}}{\Omega _{2}}\frac{2}{\eta ^{2}} 
\]

Expressing the Laguerre polynomials by their asymptotic expression $\sqrt{%
n\eta ^{2}}h_{1}\left( n;\eta ^{2}\right) $ tends, for $n\rightarrow \infty $%
, to a function depending on the product $n\eta ^{2}$ only, 
\begin{equation}
\sqrt{n\eta ^{2}}h_{1}\left( n;\eta ^{2}\right) =\tan \left( 2\sqrt{n\eta
^{2}}-\frac{\pi }{4}\right) +O\left( n^{-3/4}\right)  \label{asymptotic}
\end{equation}

Note the oscillating behavior of the eigenvalues $E\left( n\right) \sim \tan
^{2}\left( 2\sqrt{n\eta ^{2}}-\frac{\pi }{4}\right) /\eta ^{2}$. This
circumstance implies that each eigenstate is encompassed by an infinite
countable set of eigenstates of slightly different energies. For exceptional
values of $\eta $ some eigenvalues can vanish. In these cases the series
representation of the relative NCS looses its meaning. The behavior of $%
E\left( n\right) $ as $n\rightarrow \infty $ has strong implication on the
resolution of unity, as we will see in the following.

As a consequence of (\ref{asymptotic}), the logarithm of the factorial $%
\left[ h_{1}^{2}\left( n,\eta ^{2}\right) \right] !n!$ times $\eta ^{2n}$
tends asymptotically to 
\begin{eqnarray*}
&&\log \left( \left[ h_{1}^{2}\left( n;\eta ^{2}\right) \right] !n!\eta
^{2n}\right) \stackrel{n\rightarrow \infty }{\longrightarrow }\frac{\pi }{%
8\eta ^{2}}\int^{\left( \frac{8}{\pi }\right) ^{2}n\eta ^{2}}\log \left[
\tan \left( \frac{\pi }{4}\left( \sqrt{z}-1\right) \right) \right]
^{2}dz+const \\
&&\stackrel{n\rightarrow \infty }{\longrightarrow }\frac{1}{\eta ^{2}}%
u\left( \sqrt{n\eta ^{2}}\right) +const
\end{eqnarray*}
where $u\left( x\right) $ is an oscillating entire function 
\begin{eqnarray}
u\left( x\right) &=&\frac{2}{\pi }\left( -4x\Im \left[ Li_{2}\left(
e^{-i\varphi }\right) +Li_{2}\left( -e^{i\varphi }\right) \right] +\Re
\left[ Li_{3}\left( e^{-i\varphi }\right) -Li_{3}\left( -e^{i\varphi
}\right) \right] \right)  \nonumber \\
&=&\frac{4}{\pi }\sum_{k=0}^{\infty }\left( -1\right) ^{k}\frac{4x\left(
2k+1\right) -1}{\left( 2k+1\right) ^{3}}\cos \left[ \left( 2k+1\right)
4x\right]  \label{function-u}
\end{eqnarray}
with $Li_{n}\left( z\right) =\sum_{k=1}^{\infty }z^{k}/k^{n}$ the
polylogarithm function. According to (\ref{function-u}) $\log \left( \left[
h_{1}^{2}\left( n,\eta ^{2}\right) \right] !n!\eta ^{2n}\right) $ is an
oscillating function of $\sqrt{n\eta ^{2}}$, with the envelope expanding
proportionally to $\sqrt{n\eta ^{2}}$, as shown in Fig. 3. This behavior is
confirmed by the exact expressions of $\log \left( \left[ h_{1}^{2}\left(
n,\eta ^{2}\right) \right] !n!\eta ^{2n}\right) $ plotted in Fig. 4 versus $%
n $ for $\eta ^{2}=0.01,0.02,0.1$ and $0.2$.

In particular there exists a countable infinite sequence of values of $n\eta
^{2}$ for which $\left[ h_{1}^{2}\left( n,\eta ^{2}\right) \right] !n!\eta
^{2n}$ is close to $\exp \left( -\frac{16\sqrt{n}}{\pi \eta }\right) $, so
that the NCS can be normalized only for $\left| \alpha \eta \right| $ less
than one. In other words, contrarily to the linear coherent states the NCS
relative to a trapped ion fill the open circle $1/\eta $ in the complex
plane. As $\eta \rightarrow 0$ the domain of existence tends to the whole
complex plane, as for the linear coherent states. While these states are
normalized for $\alpha \eta $ inside the unit circle, the scalar product of $%
\left| \alpha _{1},h\right\rangle $ and $\left| \alpha _{2},h\right\rangle $
is defined even if one the numbers $\alpha _{1}$or $\alpha _{2}$ has an
arbitrary modulus as long as the product of the moduli is less than $1/\eta
^{2}$. A similar situation occurs for the harmonious states [22] described
by the deformation function $h\left( n\right) =1/\sqrt{n}$.

For generic combinations of vibrational excitation levels and parameter $%
\eta ^{2}$ the ion rovibronic dynamics fully displays its nonlinear
character. An example of this feature has been seen above in connection with
the discussion of the Wigner functions of some nonlinear Schr\"{o}dinger
cats and states on a circle of order 3 and 4.

\subsection{\protect\smallskip {\bf Unity resolution}}

Being these NCS restricted to values of $\alpha $ such that $\left| \alpha
\eta \right| <1$ the Mellin transform (\ref{Mellin-transform-1}) reduces to
the single-sided Laplace transform 
\[
g\left( s\right) =\lim_{\varepsilon \rightarrow 0}\int_{\varepsilon
}^{\infty }m_{h}\left( \eta ^{-2}e^{-t}\right) e^{-st}dt 
\]
Then, $g\left( s\right) $ is the right-sided Laplace transform of the
distribution $m_{h}\left( \eta ^{-2}e^{-t}\right) $ and tends asymptotically
to the analytic function $\exp \left[ \eta ^{-2}v\left( \sqrt{s\eta ^{2}}%
\right) \right] $ of $s$ in the half-plane $%
\mathop{\rm Re}%
$ $s>1$ and is bounded according to 
\[
g\left( s\right) \leq K\ e^{-T%
\mathop{\rm Re}%
s} 
\]
with $T$ a negative infinitesimally small constant and $K$ a constant.
Consequently $m_{h}$ is a distribution with support bounded on the left at $%
t=T<0$ (see Ref. [30], corollary 8.4-1a.). This means that it is necessary
to include in the unity resolution un-normalizable states of amplitude $%
\left| \alpha \right| >\eta ^{-1}$.

\subsection{\bf Approximate deformations}

The difficulties in dealing with this deformation can be overcome by using
approximate deformations. This is justified by the circumstance that in
laser cooling experiments one deals with ions occupying a finite number of
vibrational levels.

In particular when the parameter $\eta ^{2}$ is not very large the
deformations $h_{1,2}\left( n\right) $ can be approximated by a few terms of
the series expansion 
\begin{eqnarray}
h_{1}\left( \hat{n};\eta ^{2}\right) &=&1+\frac{\hat{n}-1}{2}\eta ^{2}+\frac{%
2\hat{n}^{2}-3\hat{n}+1}{6}\eta ^{4}+\frac{11\hat{n}^{3}-22\hat{n}^{2}+13%
\hat{n}-2}{48}\eta ^{6}+\cdots \   \nonumber \\
h_{2}\left( \hat{n};\eta ^{2}\right) &=&1+\frac{2}{3}\left( \hat{n}-2\right)
\eta ^{2}+\frac{11\hat{n}^{2}-39\hat{n}+34}{24}\eta ^{4}+\frac{19\hat{n}%
^{3}-96\hat{n}^{2}+159\hat{n}-86}{60}\eta ^{6}+\cdots  \label{nonlinear-ds}
\end{eqnarray}

Approximating the deformation $h_{1}\left( n\right) $ by $1+\frac{\hat{n}-1}{%
2}\eta ^{2}$ we have that $m_{\frac{\eta ^{2}}{2}h_{1}}\left( x\right) $
coincides with the Mellin antitransform of $\Gamma \left( s\right) \Gamma
^{2}\left( \frac{2}{\eta ^{2}}+s-1\right) $. Before using Eq. (\ref
{Mellin-anti}) with $A=3$, $B=0$ and $a_{1}=0$, $a_{2}=a_{3}=\frac{2}{\eta
^{2}}-1$, we have to remove the degeneracy $a_{2}=a_{3}$ by evaluating Eq.(%
\ref{Mellin-anti}) for $a_{2}-a_{3}=\varepsilon $ and letting $\varepsilon
\rightarrow 0$. Since $\lim_{\varepsilon \rightarrow 0}\Gamma \left(
\varepsilon \right) +\Gamma \left( -\varepsilon \right) \rightarrow -2\sinh
\left( \gamma \varepsilon \right) /\varepsilon =-2\gamma $ with $\gamma
=0.57721$ the Euler's constant, then 
\[
a_{2}^{2}m_{h_{1}}\left( x\right) =\Gamma ^{2}\left[ a_{2}\right] \
_{0}F_{2}\left[ ;1-a_{2,}1-a_{2};-\frac{x}{a_{2}^{2}}\right] -2\gamma \Gamma
\left[ -a_{2}\right] \Gamma \left[ 2a_{2}\right] \ _{0}F_{2}\left[ ;1,1+a;-%
\frac{x}{a_{2}^{2}}\right] \left( \frac{x}{a_{2}^{2}}\right) ^{a_{2}} 
\]

At the same time the h-exponential reads 
\[
E_{h}\left( v\right) =\ _{0}F_{3}\left[ ;1,a_{2}+1,a_{2}+1;\frac{v}{a_{2}^{2}%
}\right] 
\]

Expanding the generalized factorial $\left[ h_{1}\left( \hat{n},\eta
^{2}\right) \right] !$ to the second order in $\eta ^{2}$ the NCS $\left|
\alpha ,h_{1}\right\rangle $ can be expressed as a combination of excited
coherent states 
\[
\left| \alpha ,h_{1}\right\rangle =\left( 1+\eta ^{2}-\frac{\eta ^{4}}{12}%
\right) \left| \alpha \right\rangle +\left( \frac{\eta ^{2}}{2}+\frac{7\eta
^{4}}{72}\right) \alpha \left| \alpha ,1\right\rangle +\left( -\frac{\eta
^{2}}{2}+\frac{\eta ^{4}}{24}\right) \alpha ^{2}\left| \alpha
,2\right\rangle -\frac{5\eta ^{4}}{36}\alpha ^{3}\left| \alpha
,3\right\rangle +O\left( \eta ^{6}\right) 
\]

A better approximation can be obtained by representing the Laguerre
polynomials $L_{n-1}\left( \eta ^{2}\right) $ by a finite sum of powers of $%
\eta ^{2},$ 
\[
L_{n}\left( \eta ^{2}\right) \sim \sum_{k=0}^{K}\left( -1\right) ^{k}%
{n \choose k}%
\frac{\eta ^{2k}}{k!}=P_{K}\left( n\right) 
\]
so that $h_{1}\left( n;\eta ^{2}\right) $ (see Eq. (\ref{deformations})) can
be replaced by a rational function 
\begin{equation}
h_{A,B}\left( n\right) =\frac{P_{A+1}\left( n-1\right) -P_{A+1}\left(
n\right) }{\eta ^{2}P_{B}\left( n-1\right) }=\gamma \frac{%
\prod_{i=1}^{A}\left( n+a_{i}\right) }{\prod_{j=1}^{B}\left( n+b_{j}\right) }
\label{Pade}
\end{equation}
For using the characteristic function $\rho _{h,A}$ we should choose $A=B$,
while for introducing the P-representation $A\geq B$. In particular, Cirac
et al [8][13] and Blockley et al [9] have expanded the exponential of Eq. (%
\ref{A}) up to the second order in $\eta $, their case corresponds to $A=B=4$%
.

If the roots $-a_{i}$ and $-b_{i}$ are not integer we have (cf. Eq. (\ref
{fact-gen})) 
\[
\left[ h_{AB}\left( s\right) \right] !=\gamma ^{s-1}\Gamma ^{-1}\left[ 
{\left( a\right)  \atop \left( b\right) }%
\right] \Gamma \left[ 
{\left( a\right) +s \atop \left( b\right) +s}%
\right] 
\]
For this class of deformations the measure is given by the combination of
the generalized hypergeometric functions of Eq. (\ref{measure-scaling}),
subject to the precaution of removing the degeneracy of the coefficients $%
a_{i},b_{i}$.

In particular for $A=B=1$ the Husimi--Kano Q-function (\ref{Husimi-Kano})
relative to the state $\left| z,h\right\rangle $ reduces to 
\[
Q_{1}(w)\varpropto e^{-\left| w-z\right| ^{2}/2}\left| \varpi ^{\nu }J_{-\nu
}\left( \varpi \right) \right| ^{2}\longrightarrow _{\left| \varpi \right|
\rightarrow \infty }Ce^{-\left| w-z\right| ^{2}/2}\left[ 1+\frac{\cot \left(
\nu \pi \right) }{2}\left( \zeta w^{*2\nu }+\zeta ^{*}w^{2\nu }\right)
\right] 
\]
with $\nu =\frac{1}{\eta ^{2}}+\frac{1}{2}$, $\varpi =\frac{i}{2}zw^{*}$ and 
$\zeta =\left( i\frac{ez}{4\nu }\right) ^{2\nu }$. Consequently, the
projector $\left| z,h\right\rangle \left\langle z,h\right| $ is represented
in terms of the undeformed coherent states by a P-representation containing
derivatives of the Dirac function of the very high order $\nu $. This
confirms the advantage of using the representation (\ref{P-representation}).

For a finite rank density matrix $m_{h}\left( x\right) $ can be represented
by a finite combination of Laguerre polynomials 
\begin{equation}
m_{h}\left( x\right) =e^{-x}\sum_{n=0}^{n_{\max }}m_{n}L_{n}\left( x\right)
\label{measure-Laguerre}
\end{equation}
Imposing the condition (\ref{integral-equation-mu}) for $0\leq n\leq n_{\max
}$ yields 
\[
m_{n}=\sum_{m}\left( -1\right) ^{m}%
{n \choose m}%
\left( \left[ h\left( m\right) \right] !\right) ^{2} 
\]
In Fig. 5 we have plotted these approximate measure functions for different
values of $\eta ^{2}\left( =0.015,0.0156,0.0158,0.016\right) $, representing
density operators relative to ions excited up to the level $n=50$. For these
values of $\eta ^{2}$ the inclusion of a larger number of terms $\left(
n>50\right) $ would lead to measures taking negative values.

\section{\bf Conclusions}

The vibrational steady states of ions placed in a parabolic trap and driven
by bichromatic fields detuned by multiples of the vibrational frequency
provide a class of realizations of the nonlinear version of the so-called
coherent states on a circle. The most well known example is that of the
nonlinear Schr\"{o}dinger cat states. As for the linear case also these
states can be decomposed into finite sums of nonlinear coherent states,
which can be considered as the building blocks of the vibrational
wavefunctions of systems driven by laser fields detuned by multiples of the
vibrational frequency.

This class of states is well described by the Wigner function, which has
been computed for states on a circle of degree two (cats), three and four.
The relative patterns show a dramatic dependence on the Lamb-Dicke parameter 
$\eta $, which measures the degree of departure from the linear case. This
behavior is due to the irregular dependence of the ion deformation function
on the Fock state index n.

With the aim of investigating the possibility of extending some mathematical
tools of the linear coherent state theory to NCS it has been introduced a
deformed displacement operator ${\cal D}_{h}\left( \alpha \right) $, which
in analogy with the linear one generates the NCS $\left| \alpha
,h\right\rangle $ by displacing the Fock vacuum state $\left( \left| \alpha
,h\right\rangle ={\cal D}_{h}\left( \alpha \right) \left| 0\right\rangle
\right) $. The penalty paid for this extension is the loss of the unitarity.
However, it allows the construction of a linear functional, which can be
used for representing density operators by means of a generalized
probability distribution function $\rho _{A,h}\left( z\right) $.

The peculiarities of NCS connected with trapped ions become evident when the
deformation factorial is analyzed for very large $n$. The weight of the n-th
Fock state contributing to a NCS exhibits an almost periodic behavior by
taking very large and very small values of the order of $e^{\pm C\sqrt{n\eta
^{2}}}$. Consequently, the NCS can be normalized only for $\alpha $ filling
the open circle $1/\eta $ in the complex $\alpha $-plane. In addition, this
behavior prevents the existence of a regular measure for resolving the
unity. These pathologies mark the difference with q-oscillators whose
deformation function is an increasing function of $n$.

For extending to these NCS the P-representation formalism it is necessary to
replace the deformation with an approximate one, for which there exists a
measure. These approximate NCS can be constructed in different ways. Two
examples are provided. In the first case the deformation is represented by a
rational function of the occupation number, obtained by truncating the
Laguerre polynomials to some order in $\eta ^{2}$. By increasing the degree
of these rational functions it is possible to represent accurately the
actual deformation for occupation numbers extending up to infinity. The
respective measures are the Mellin antitransforms of Gamma function products 
$\Gamma \left[ 
{\left( a\right)  \atop \left( b\right) }%
\right] $ and are given by combinations of generalized hypergeometric
functions. In the other case the measure is represented by a finite
combination of Laguerre polynomials. In this way it is possible to represent
exactly the factorials up to a given level $n$, although it is not possible
to obtain in general a positive definite measure.

The NCS can provide a basis for studying the trapped ion evolution by
representing the statistical expectation values of either antinormal $%
G_{A}\left( A,A_{h}^{\dagger }\right) $ or normal $G_{N}\left(
A,A_{h}^{\dagger }\right) $products of $A$ and $A_{h}^{\dagger }$ as
integrals of the probability distributions $\rho _{h,A}\left( z\right) $ or $%
P_{h}\left( z\right) $ times the classical functions $G_{A}\left(
z,z^{*}\right) $ and $G_{N}\left( z,z^{*}\right) $. Another possibility
consists in transforming the density matrix master equation in an equivalent
equation for the P-representation. This problem will be addressed in a more
systematic way in a forthcoming paper.

Before concluding, it is remarkable that the deformation used for the
q-oscillators, the ancestors of the NCS, is based on the same transformation
used by Heine [31] a century ago for generalizing the Gauss hypergeometric
function.

\section*{Acknowledgments}

V.I.M. thanks the University of Naples ``Federico~II'' for the kind
hospitality and the Russian Foundation for Basic Research for the partial
support under Project~No.~99-02-17753.

\smallskip

\begin{center}
\smallskip {\bf References}
\end{center}

[1] L. Biedenharn, J. Phys. {\bf A22, }L873 (1989)

[2] A. MacFarlane, J. Phys. {\bf A22, }4581 (1989); {\it see also }C. P. Sun
and H.-C Fu, J. Phys. {\bf A22, }L983 (1989); M. Chaichian and P. Kulish,
Phys. Lett. {\bf B232, }72 (1990)

\smallskip [3] R. W. Gray and C. A. Nelson, J. Phys. {\bf A23, L945} (1990);
A. J. Bracken, D. S. McAnally, R. B. Zhang and M. D. Gould, J. Phys. {\bf %
A24, }1379 (1991); B. Jurco, Lett. Math. Phys. {\bf 21, }51 (1991); C. A.
Nelson, ''Novel implications of the q-analogue coherent states'', in
''Symmetries in Science VI'', ed. B. Gruber, Plenum Press, N.Y., p.563,
(1993).

[4] V. I. Man'ko, G. Marmo, E.C.G. Sudarshan, and F. Zaccaria, Physica
Scripta {\bf 55, }528 (1997); V. I. Man'ko G.Marmo, and F. Zaccaria in
''Symmetries in Science'', Ed. B. Gruber and M. Ramek, Plenum Press, N. Y.
(1997);

[5] J. \v {C}rnugelj, M. Martinis, and V. Mikuta-Martinis, Phys. Lett. {\bf %
A 188}, 347 (1994); Phys. Lett. {\bf B 318}, 227 (1993); Phys. Rev. {\bf A 50%
}, 1785 (1994);

\smallskip [6] E. T. Jaynes and F. W. Cummings, Proc. IEEE {\bf 51}, 89
(1963);

[7] B. Buck and C. V. Sukumar, Phys. Lett. {\bf 81 A}, 132 (1981); J. Phys. 
{\bf A 17}, 885 (1984); V. Buzek, Phys. Rev. {\bf A 39}, 3196 (1989), G. S.
Agarwal, J. Opt. Soc. Am. {\bf B2}, 480 (1985); C. C. Gerry, Phys. Rev. {\bf %
A 37}, 2683 (1988); C. V. Sukumar and B. Buck, Phys. Lett. {\bf 83 A}, 211
(1981); A. S. Shumovsky, Fam Le Kien and E. I. Aliskenderov, Phys. Lett. 
{\bf A 124}, 351 (1987);

[8] J. J. Cirac, R. Blatt, P. Zoller, and W. D. Phillips, Phys. Rev. {\bf A
46,} 2668 (1992)

[9] C. A. Blockley and D. F. Walls, Phys. Rev. {\bf A 47}, 2115 (1993); C.
A. Blockey, D. F. Walls, and H. Risken, Europhys. Lett. {\bf 17}, 509 (1992)

\smallskip [10] J. H. Eberly, N. B. Narozhny, and J. J. Sanchez-Mondragon,
Phys. Rev. Lett. {\bf 44}, 1323 (1980); N. B. Narozhny, J. J. Sanchez, and
J. H. Eberly, Phys. Rev. {\bf A 23}, 236 (1981); P. Meystre and M. S.
Zubairy, Phys. Lett. {\bf 89 A}, 390 (1982); C. C. Gerry, Phys. Rev. {\bf A
37}, 2683 (1988); J. R. Kukeinski and J. L. Madajczyk, Phys. Rev. {\bf A 37}%
, 317 (1988);

[11] F. Diedrich and H. Walter, Phys. Rev. Lett. {\bf 58}, 203 (1987); M.
Schubert, I. Siemers, R. Blatt, W. Neuhauser, and P. E. Toscheck, Phys. Rev.
Lett. {\bf 68}, 3016 (1992)

[12] W. Vogel, J. Phys. {\bf B}: At. Mol. Phys. {\bf 16,} 4481 (1983); W.
Vogel and Th. Ullmann, J. Opt. Soc. Am. {\bf B3}, 441 (1986)

[13] J. J. Cirac, R. Blatt, A. S. Parkins, and P. Zoller, Phys. Rev. Lett. 
{\bf 70}, 556 (1993); J. I. Cirac, P. Zoller, Phys. Rev. Lett. 74, 4091
(1995).

[14] R. L. de Matos Filho and W. Vogel, Phys. Rev. {\bf A49}, 2812 (1994);
Phys. Rev. {\bf A 54}, 4560 (1996);

[15] S. Mancini, Phys. Lett. {\bf A 233, }291 (1997).

[16] X-G Wang and H-C Fu, quant-ph/9903013 vs. Nov (1999).

[17] V. Bargmann, Commun. Pure and Appl. Math. {\bf 14,} 187 (1961).

[18] C. L. Mehta and E.C.G. Sudarshan, Phys. Rev. {\bf 138B}, 274 (1965);
see also J. R. Klauder, J. Math. Phys. {\bf 5}, 177 (1964) and T. F. Jordan,
Phys. Lett. {\bf 11}, 289 (1964)

[19] F. Diedrich, J. C. Bergquist, Wayne M. Itano, and D. J. Wineland, Phys.
Rev. Lett. {\bf 62,} 403 (1989).

[20] G.S. Agarwal and K. Tara, Phys. Rev. {\bf A 43, }492 (1991); V. V.
Dodonov, Ya. A. Korennoy, and V. I. Man'ko and Y.A. Moukhin, Quantum and
Semiclassical Optics {\bf 8,} 413 (1996);

[21] S. Sivakumar, quant-ph/9806061;

[22] R. J. Glauber, Phys. Rev. Lett., {\bf 10}, 84 (1965); E. C. G.
Sudarshan, Phys. Rev. Lett. {\bf 10}, 277 (1965);

[23] E. C. G. Sudarshan, Int. J. Theor. Phys. {\bf 32}, 1069 (1993)

[24 ] L. J. Slater, ''Generalized Hypergeometric Functions'', Cambridge
University Press, Cambridge (1966); Proc. Camb. Phil. Soc. {\bf 51}, 577
(1955)

[25] Jinzuo Sun, Jisuo Wang, and Chuankui Wang, Phys. Rev. {\bf A 44}, 3369
(1991);

[26] O. Casta\~{n}os, R. Lop\'{e}z-Pe\~{n}a, V.I. Man'ko, J.Russ. Laser
Res., {\bf 16}, 477 (1995)

[27] J. Janszky, P. Domokos, and P. Adam, Phys. Rev. {\bf A 48}, 2213
(1994); Ren-Shan Gong, Phys. Lett. {\bf A 233,} 297 (1997).

[28] E. Wigner, Phys. Rev., {\bf 40}, 749 (1932); K. Husimi, Proc. Phys.
Math. Soc (1965).

[29] B. Roy, Phys. Lett. A {\bf 249,} 25 (1998); B.Roy and P. Roy, Phys.
Lett. A {\bf 263,} 48 (1999)

[30] A. H. Zemanian, ''Distribution Theory and Transform Analysis'',
McGraw-Hill, New York (1965).

[31] E. Heine, Handuch die Kugelfunctionen, (1898) quoted by Slater
(Ref.[24]).

\begin{center}
{\bf Figure captions}
\end{center}

Fig. 1: Top and side views of the Wigner function (Eq. \ref{Wigner})
relative to linear (a) and nonlinear (b,c,d) even Schr\"{o}dinger cat states
for different values of $\alpha =3.5$ (real) and two different values of $%
\eta $. (a) $\eta =0$ (linear case), the two coherent peaks are almost
circularly shaped the interference pattern is axially symmetric along the
axis defined by the center of the coherent peaks; (b -- top-view, c --
side-view and d -- air-view) $\eta =0.5$, the gaussian peaks are no more
distinguishable from the circularly symmetric interference pattern.

Fig. 2: Top view of the Wigner function (Eq. \ref{Wigner}) for 3 and 4 NCS
states sitting on the circle. $\alpha =3.5$ (real) and $\eta =0.33$. This
value of the nonlinearity $\eta $ leads to a smooth reshaping of the
coherent peaks from a nearly circular shape to an elliptic one.

Fig. 3 Asymptotic expression of $u\left( \sqrt{z}\right) $ versus $z$ (see
Eq. (\ref{function-u}))

Fig. 4 $\log \left( \left[ h_{1}^{2}\left( n,\eta ^{2}\right) \right]
!n!\eta ^{2n}\right) $ versus $n$ for $\eta ^{2}=0.01$ (a),$0.02$ (b)$,0.1$
(c) and $0.2$ (d).

Fig. 5 Measure $m_{h}\left( x\right) $ versus x obtained as a combination of
50 Laguerre polynomials (Eq. (\ref{measure-Laguerre})) and four different
values of $\eta ^{2}=0.015,0.0156,0.0158,0$.$016$

\end{document}